\documentclass[%
 reprint,
 amsmath,amssymb,
 prb,
]{revtex4-1}

\usepackage{graphicx}
\usepackage{dcolumn}
\usepackage{bm}

\usepackage{mathtools}

\usepackage{siunitx}
\usepackage{multirow}
\usepackage{array}
\newcolumntype{L}[1]{>{\raggedright\let\newline\\\arraybackslash\hspace{0pt}}m{#1}}
\newcolumntype{C}[1]{>{\centering\let\newline\\\arraybackslash\hspace{0pt}}m{#1}}
\newcolumntype{R}[1]{>{\raggedleft\let\newline\\\arraybackslash\hspace{0pt}}m{#1}}

\begin{document}

\title{Self-compensation due to point defects in Mg-doped GaN}

\author{Giacomo Miceli}
\email{giacomo.miceli@epfl.ch}
\author{Alfredo Pasquarello}
\affiliation{Chaire de Simulation \`a l'Echelle Atomique (CSEA), %
             Ecole Polytechnique F\'ed\'erale de Lausanne (EPFL),
             CH-1015 Lausanne, Switzerland} 
             
\date{July, 8 2015}

%
%
\begin{abstract}
Using hybrid density functional theory, we address point defects susceptible to 
cause charge compensation upon Mg doping of GaN. We determine the free-energy of
formation of the nitrogen vacancy and of several Mg-related defects. The entropic
contribution as a function of temperature is determined within the quasiharmonic
approximation. We find that the Mg interstitial shows a noticeably lower free-energy
of formation than the Mg substitutional to Ga in $p$-type conditions. Therefore,
the Mg impurity is amphoteric behaving like an acceptor when substitutional to 
Ga and like a double donor when accommodated in an interstitial position. The 
hybrid-functional results are then linked to experimental observations by solving 
the charge neutrality equations for semiconductor dominated by impurities. We show 
that a thermodynamic equilibrium model is unable to account for the experimental 
hole concentration as a function of Mg doping density, due to nitrogen vacancies 
and Mg interstitials acting as compensating donors. To explain the experimental 
result, which includes a drop-off of the hole concentration at high Mg densities, 
we thus resort to non-equilibrium models. We show that either nitrogen vacancies 
or Mg interstitials could be at the origin of the self-compensation mechanism. 
However, only the model based on interstitial Mg donors provides a natural mechanism 
to account for the sudden appearance of self-compensation. Indeed, the amphoteric
nature of the Mg impurity leads to Fermi-level pinning and accounts for the observed
drop-off of the hole concentration of GaN samples at high Mg doping. Our work 
suggests that current limitations in $p$-type doping of GaN could be overcome by
extrinsically controlling the Fermi energy during growth. 
\end{abstract}

\maketitle

%
%
\section{Introduction}
\label{sec:intro}
Gallium nitride is already an essential compound for commercial blue-light-emitting 
diodes and represents a very promising material for future device applications.%
\cite{pearton_JAP1999} Further progress in this field requires achieving high 
concentrations of free carriers in both $n$-type and $p$-type layers. However, 
the $p$-type doping efficiency in GaN is still too low and is one of the major 
problems hampering the widespread use of this material in optoelectronic applications.
The Mg impurity substituting Ga, Mg$_\text{Ga}$, has hitherto been recognized 
as the only effective acceptor source in GaN.\cite{amano_JJAP1989,nakamura_APL1994}
Due to its ionization energy of about 220 meV,
\cite{madelung2004,monemar_PRL2009,monemar_PSSC2010,monemar_JAP2014} high Mg doping
levels are needed in order to achieve significant hole concentrations at room 
temperature. However, the doping efficiency breaks down at high Mg densities, 
thereby limiting the hole concentrations that can be achieved in practice.
\cite{kaufmann_PRB2000,brochen_APL2013,malinverni_APL2014}

At present, metallorganic vapor phase epitaxy (MOVPE) is the most used technique
to grow $p$-type GaN:Mg layers. However, the doping efficiency of as-grown samples
is extremely low, due to hydrogen playing a critical passivation role. The hydrogen
passivation effect during the growth of GaN:Mg layers has been considered as a 
beneficial effect. Indeed, the passivation of substitutional Mg impurities keeps
the Fermi-level high in the band gap, thereby preventing the formation of compensating
donors.\cite{neugebauer_APL1996} The acceptors are then activated through post-growth
annealing treatments.\cite{amano_JJAP1989,nakamura_JJAP1992} For samples grown by
metallorganic chemical vapor deposition, Kaufmann {\it et al.}\cite{kaufmann_PRB2000}
have shown that hydrogen depassivation is very effective in the moderate doping 
range, i.e.\ for Mg densities raging between $3\times10^{18}$ and $2\times10^{19}$
cm$^{-3}$. However, more recently, it has been shown that hydrogen incorporates 
proportionally with Mg, likely forming beneficial Mg-H complexes, but saturates 
at a Mg doping threshold of about 3$\times$10$^{19}$ cm$^{-3}$.\cite{castiglia_APL2011}
Therefore, higher Mg doping densities cannot be achieved through the beneficial 
passivation effect of hydrogen. In addition, post-growth annealing treatments are
not able to entirely remove the hydrogen atoms.\cite{castiglia_APL2011} As a 
consequence, part of the Mg$_\text{Ga}$ acceptors are passivated by hydrogen and
remain electrically inactive. On the basis of these considerations, the use of 
hydrogen as a temporary passivating agent during growth does not allow one to 
envisage higher hole concentrations in GaN.

Molecular beam epitaxy (MBE) has been proposed as an alternative growth technique
to overcome the doping limitations described above. In this growth technique, the
lower operating temperatures enable higher hole densities, in the absence of any
significant hydrogen concentration.\cite{namkoong_APL2008,brochen_APL2013,malinverni_APL2014}
Nevertheless, regardless of the adopted growth technique, all the experimental 
studies report a drastic drop-off in the hole density upon reaching typical Mg 
concentrations of about $10^{19}$ cm$^{-3}$.\cite{castiglia_APL2011,kaufmann_PRB2000,brochen_APL2013} 
Hence, we conclude that the origin of the drop-off should not be related to the 
occurrence of hydrogen.

The drastic decrease in hole concentration above a Mg doping density of $10^{19}$
cm$^{-3}$ could arise from a deterioration of the sample. Above the solubility 
limit, the excess of Mg would precipitate forming clusters of new phases causing 
the degradation of the crystallinity. Such a deterioration in semiconductor samples 
would dramatically affect the carrier transport properties. However, the measured 
hole mobility in GaN:Mg does not undergo any dramatic change over the doping range 
$3\times10^{18}$--$7\times10^{19}$ cm$^{-3}$.\cite{kaufmann_PRB2000}

The experimental evidence mentioned above suggests that neither the presence of 
hydrogen nor the crystalline deterioration can cause the drop-off in the hole 
concentration at the Mg doping density of about $10^{19}$ cm$^{-3}$. Hence, a 
compensation mechanism based on either intrinsic or Mg-related point defects is 
generally invoked. 

In early studies, the key role in the self-compensation process has generally 
been assigned to the nitrogen vacancy, V$_\text{N}$.%
\cite{kaufmann_APL1998,kozodoy_JCG1998,vandewalle_JCG1998} Later, the 
Mg$_\text{Ga}$-V$_\text{N}$ defect complex, a deep donor defect in GaN, has 
been associated with the observed photoluminescence (PL) peak at about 2.9 eV
and assumed to be at the origin of the severe compensation in heavily doped 
GaN:Mg.\cite{kaufmann_PRB2000,kozodoy_JAP2000} The Mg$_\text{Ga}$-V$_\text{N}$
complex has indeed been identified through positron annihilation spectroscopy.%
\cite{hautakangas_PRL2003,hautakangas_PRB2005} However, this defect complex turned
out to occur only in moderate concentrations ($\sim$$2 \times 10^{17}$ cm$^{-3}$)
and to be unstable against annealing above 500$^{\circ}\text{C}$.%
\cite{hautakangas_PRL2003,hautakangas_PRB2005} This experimental evidence clearly
contrasts with the dominance of the peak at 2.9 eV in the measured PL spectra.%
\cite{monemar_PSSC2010} The role of the Mg$_\text{Ga}$-V$_\text{N}$ complex  
has further been diminished by a theoretical study, \cite{lyons_PRL2012} in which
the blue luminescence has instead been associated to the substitutional Mg$_\text{Ga}$
impurity. In a recent theoretical work,\cite{buckeridge_PRL2015} the V$_\text{N}$
defect has been found at noticeably lower energies than beforehand,\cite{yan_APL2012}
reviving the suggestion that this defect plays a primary role in the compensation.
Hence, despite the importance of the technological implications and the numerous
efforts devoted to this problem, the microscopic mechanisms behind the 
self-compensation process in $p$-type GaN:Mg have remained elusive to a large 
extent.

In this work, we address the self-compensation mechanism in GaN by an extensive 
investigation of the role played by point defects upon Mg doping. Through hybrid
density functional calculations, we first obtain the formation energies of a set
of relevant point defects, including the nitrogen vacancy (V$_\text{N}$), the 
magnesium substitutional to gallium (Mg$_\text{Ga}$), the magnesium interstitial 
(Mg$_\text{inter}$), and the Mg$_\text{Ga}$-V$_\text{N}$ defect complex.
Among the native donor point defects, the nitrogen vacancy is found to be the most 
stable defect, in agreement with previous theoretical studies.\cite{yan_APL2012,miceli_MEE2015}
Furthermore, we find that the Mg impurity in GaN shows an amphoteric behavior 
acting like an acceptor in the form of Mg$_\text{Ga}$ and like a double donor 
in the form of Mg$_\text{inter}$, when going from $n$-type to $p$-type conditions. 
Through the equations of semiconductors dominated by impurities, we link our 
hybrid-functional results with experimental observations. The amphoteric nature 
of the Mg impurity is identified as the mechanism causing Fermi-level pinning and as  
origin of the observed drop-off of hole concentrations in GaN samples with 
increasing Mg doping.

The article is organized as follows. In Sec.\ \ref{sec:thermodynamics}, the 
relevant point defects are studied within a hybrid-functional framework. In 
particular, in Sec.\ \ref{subsec:free-energy}, we describe the theoretical 
formulation for determining the energetics of point defects. The defect formation
energies and charge transition levels are given in Sec.\ \ref{subsec:defects}. 
In Sec.\ \ref{sec:self}, we use the charge-neutrality equations of semiconductors
dominated by impurities to investigate the hole concentration as a function of 
Mg doping density. We focus on equilibrium and non-equilibrium models in Secs.\ 
\ref{subsec:equilibrium} and \ref{subsec:non-equilibrium}, respectively. Conclusions
are drawn in Sec.\ \ref{sec:conclusions}.

\section{Thermodynamics of point defects}
\label{sec:thermodynamics}


\subsection{Formation free energy}

\label{subsec:free-energy}
In this work, we focus on the defect formation free energy which is a key thermodynamic 
quantity determining the defect concentration. When a point defect $X$ is formed
in GaN in the charge state $q$ and at a given temperature $T$, the required 
free energy of formation is defined as follows 
\begin{align}
F^\text{f}({X}^q;T)  = & F_\text{tot}({X}^q;T) - F_\text{tot}(\text{GaN};T) %
  - \sum_\alpha n_\alpha \mu_\alpha(T) \nonumber \\
  & + q \left( \varepsilon_\text{v} + \mu_\text{F} + \Delta V \right) + E^q_\text{corr},
\label{eq:freeener1}
\end{align}
where $F_\text{tot}({X}^q;T)$ and $F_\text{tot}(\text{GaN};T)$ represent the 
calculated total free energies for the defective and pristine supercells, respectively. 
$n_\alpha$ is the number of atoms of the species $\alpha$ added ($n_\alpha > 0$) 
or removed ($n_\alpha < 0$) from the system and $\mu_\alpha(T)$ represents
its relative chemical potential. $\mu_\text{F}$ is the electronic chemical
potential as referred to the top of the valence band $\varepsilon_\text{v}$. $\Delta V$ is an 
alignment term and $E^q_\text{corr}$ is the correction term due to finite-size 
effects.\cite{freysoldt_PRL2009,komsa_PRB2012}

\begin{figure}
\centering
\includegraphics[width=8cm]{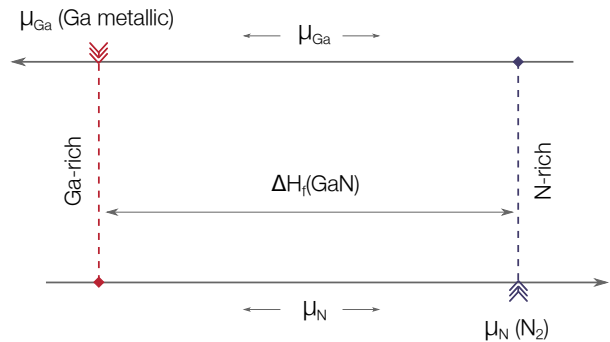}
\caption{Schematic representation of the thermodynamically allowed experimental 
         conditions for the growth of GaN. Ga and N chemical potentials can vary
         within an energy range set by the enthalpy of formation of GaN. The 
         extreme cases are referred to as Ga-rich and N-rich conditions, 
         respectively.}
\label{fig:therm}
\end{figure}

The defect formation free energy depends on the relative abundances of the atomic
species involved in the growth process, which are set by the choice of the reference 
elemental chemical potentials. In the case of GaN, the chemical potentials 
$\mu_\text{N}$ and $\mu_\text{Ga}$ are linked by the equilibrium thermodynamic
condition which guarantees the stability of the GaN phase:
\begin{equation}
\mu_\text{GaN}(T) = \mu_\text{Ga}(T) + \mu_\text{N}(T).
\label{eq:stability}
\end{equation}
Depending on the experimental growth conditions, $\mu_\text{N}(T)$ and 
$\mu_\text{Ga}(T)$ can vary within specific bounds. For Ga and N 
chemical potentials, the upper bounds are set by the formation of metallic Ga 
and of the nitrogen molecule, respectively:
\begin{align}
 \mu_\text{Ga}(T) & \le \mu_\text{Ga}(\text{Ga metallic};T) \\
 \mu_\text{N}(T)  & \le \mu_\text{N}(\text{N}_2;T).
\end{align}
The following condition holds between the chemical potentials in the extreme 
cases: 
\begin{eqnarray}
\mu_\text{GaN}(T) &=& \mu_\text{Ga}(\text{Ga metallic};T) + \mu_\text{N}(\text{N}_2;T)
\nonumber \\
&& + \Delta \text{H}_\text{f}(\text{GaN};T)
\label{eq:stability2}
\end{eqnarray}
where $\Delta \text{H}_\text{f}(\text{GaN};T)$ is the formation enthalpy of
the GaN compound. Hence, the formation enthalpy of GaN sets the range of variation
for the elemental chemical potentials (cf.\ Fig.\ \ref{fig:therm}). In Fig.\ \ref{fig:therm},
we schematically represent the range of variation of the elemental chemical 
potentials which fixes the equilibrium properties of the sample. Any possible 
thermodynamic condition for growing GaN is represented by a vertical line located
between the two thermodynamic extremes corresponding to Ga-rich and N-rich conditions. 

As far as magnesium is concerned, we set the chemical potential to a value 
corresponding to the equilibrium of the compound Mg$_3$N$_2$, which is the most 
stable Mg compound at fixed $\mu_\text{N}$.\cite{neugebauer_MRSP1995} Hence, 
once $\mu_\text{N}$ and $\mu_\text{Ga}$ are set, the Mg chemical potential 
is obtained from the following equilibrium condition: 
\begin{equation}
3\mu_\text{Mg}(T) + 2\mu_\text{N}(T) = \mu_{\text{Mg}_3\text{N}_2}(T).
\end{equation}

%
Vibrational contributions are rarely considered in density-functional studies 
of defects in solid-state systems,%
\cite{Al-Mushadani_PRB2003,estreicher_PRB2004,grabowski_JPSB2011,miceli_PRB2011}
as they are generally negligible at room temperature. However, in this work, we 
need to determine defect concentrations at growth temperatures as high as 1300 K,
and finite temperature effects might play an important role. For this reason, we
here consider vibrational contributions within the quasiharmonic approximation. 
In this approximation, the vibrational free energy is expressed in terms of the 
harmonic frequencies $\omega_i$  at zero temperature: \cite{chandler1987}
\begin{equation}
F_\text{vib}(T) = \sum_i \frac{\hbar \omega_i}{2} +
             \sum_i k_\text{B}T \ln \left[ 1- \exp \left( - \frac{\hbar \omega_i}{k_\text{B}T } \right) \right],
\label{eq:qhapprox}
\end{equation}
where the first sum is the zero-point energy and the second sum corresponds to 
the temperature-dependent entropic contributions. Anharmonic effects are neglected
in this approximation. To highlight the corrections going beyond the formation-energy
formulation at zero temperature, we rewrite Eq.\ (\ref{eq:freeener1}) as
\begin{eqnarray}
F^\text{f}({X}^q;T) & \cong & E_\text{tot}({X}^q) - E_\text{tot}(\text{GaN}) %
  - \sum_\alpha n_\alpha \mu_\alpha^0  \nonumber \\
  && + q \left( E_v + \mu_\text{F} + \Delta V \right) + E^q_\text{corr} 
  + \Delta E_\text{QH}(T), \nonumber \\
\label{eq:freeener}
\end{eqnarray}
where $E_\text{tot}({X}^q)$ and $E_\text{tot}(\text{GaN})$ are the zero-temperature
total energies for the defective and pristine supercells, respectively, and 
$\mu_\alpha^0$ is the zero-temperature chemical potential of the species $\alpha$.
The zero-point internal energy and the vibrational contributions resulting from 
the quasiharmonic approximation are included in the term $\Delta E_\textrm{QH}(T)$,
which is defined as
\begin{eqnarray}
  \Delta E_\text{QH}(T) & = & F_\text{vib}(\text{X}^q;T) - F_\text{vib}(\text{GaN};T)  \nonumber \\
  & & - \sum_\alpha n_\alpha \left[ \mu_\alpha(T) - \mu_\alpha^0 \right].
\label{eq:qh}
\end{eqnarray}

\subsection{Relevant point defects}
\label{subsec:defects}

The calculations in this work are performed with the hybrid density functional 
proposed by Heyd, Scuseria, and Ernzerhof (HSE).\cite{hse,*hse_erratum} We include
a fraction of Fock exchange equal to 31\% to reproduce the experimental band gap
of GaN. Structural properties are not significantly influenced by the adopted 
fraction of Fock exchange.\cite{chen_PRB2014} Our computational scheme relies on
norm-conserving pseudopotentials and plane-wave basis sets, as made available in the 
{\sc Q}uantum-\textsc{Espresso} suite of programs.\cite{quantum_espresso} We use
the HSE implementation described in Ref.\ \onlinecite{komsa_PRB2010}. The kinetic
energy cutoff for the wave functions is set at 45 Ry. Spin-unrestricted calculations
are performed whenever unpaired electrons occur. Defects are modeled starting from
a pristine bulk supercell containing 96 atoms and all defect structures are fully
relaxed at the hybrid functional level. In the relaxation, the Brillouin zone of
the supercell is sampled at the $\Gamma$ point and the exchange potential is 
treated as described in Ref.\ \onlinecite{broqvist_PRB2009}.

The energetics and the electronic structure of the optimized defect geometries 
are then evaluated with a finer $2\times2\times2$ Monkhorst-Pack grid in the Brillouin
zone of the supercell. We verified the accuracy of this scheme performing calculations
with the semilocal density functional proposed by Perdew, Burke, and Ernzerhof 
(PBE).\cite{pbe} We focus on the nitrogen vacancy defect for which structural 
relaxations are most critical. The nitrogen vacancy in its charge state +1 is 
fully relaxed using both a $\Gamma$-point and a $2\times2\times2$ mesh of {\bf k}
points. The resulting formation energies are found to differ by less then 1 meV.

Throughout this work, we do not explicitly include gallium $3d$ electrons among the 
valence states following previous theoretical studies on nitrides.\cite{gordon_PRB2014} 
To validate this approximation, we perform PBE calculations with and without 3$d$
states in the valence. The formation energy of the nitrogen vacancy in its charge
state +1 is found to be affected by less than 0.1 eV.

The supercell approach to determine the energetics of charged defects requires 
some specific care due to the occurrence of spurious electrostatic interactions.
For an accurate description of the energetics of isolated charged defects we apply
state-of-the-art finite-size corrections.\cite{freysoldt_PRL2009,komsa_PRB2012}
We note that we will adopt the isolated approximation also for defect densities 
as high as $3.5 \times 10^{19}$ cm$^{-3}$. To estimate possible errors, we consider
the finite-size energy correction pertaining to these high defect concentrations.
In the case of $q=\pm 1$ charge states, as for Mg$_\text{Ga}^-$ and V$_\text{N}^+$, 
we obtain an energy correction of 0.075 eV. For $q=2$, as for Mg$_\text{inter}^{+2}$, 
the correction yields 0.30 eV. In particular, these corrections would affect the
charge transition level between the Mg$_\text{Ga}^-$ and Mg$_\text{inter}^{+2}$ 
states by only 0.075 eV. These errors are sufficiently small to be neglected 
(cf.\ discussion in Sec.\ \ref{subsub:mg}).

%

\begin{figure}
\centering
\includegraphics[width=8cm]{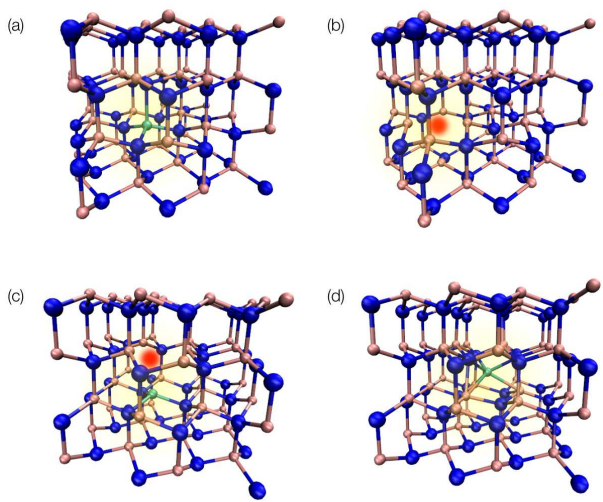}
\caption{Atomistic models of various relevant point defects in GaN: 
         (a) the Mg impurity substitutional to Ga (Mg$_\text{Ga}$), 
         (b) the nitrogen vacancy (V$_\text{N}$), 
         (c) the Mg$_\text{Ga}$-V$_\text{N}$ complex, and 
         (d) the Mg interstitial. The Mg impurity is shown in green,
         while the vacancy site is indicated by a blurred red sphere. }
\label{fig:cell}
\end{figure}

\begin{figure}
\centering
\includegraphics[width=8cm]{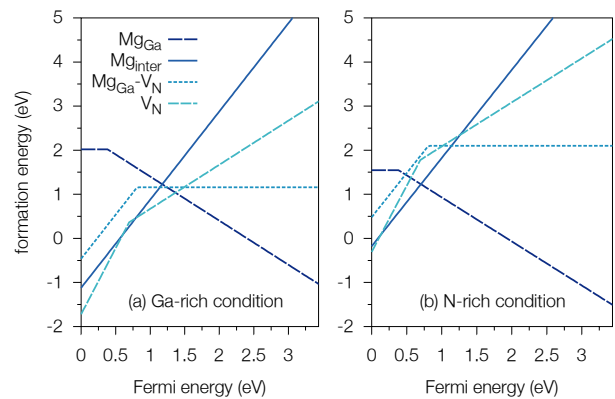}
\caption{Calculated formation energies vs Fermi energy for various relevant 
         point defects in GaN: Mg$_\text{Ga}$, Mg$_\text{inter}$, 
         V$_\text{N}$, and the Mg$_\text{Ga}$-V$_\text{N}$ complex, 
         in (a) Ga-rich and (b) N-rich conditions.}
\label{fig:energies}
\end{figure}

\begin{table*}
\label{tab:energies}
\caption{Calculated defect formation energies ($E_\text{f}^q$) and charge 
         transition levels ($\varepsilon_{q/q'}$) for relevant charge states 
         ($q$ and $q^\prime$), compared to previous theoretical work. The 
         Fermi-energy is fixed at the VBM. Energies are in eV. Some of the
         values taken from Refs.\ \onlinecite{lyons_PRL2012,yan_APL2012,buckeridge_PRL2015}
         are inferred from figures.}
\begin{ruledtabular}
\begin{tabular}{llddddd}
  &  &  \multicolumn{2}{c}{Ga-rich} & \multicolumn{3}{c}{N-rich} \\
  Defect &%
   &%
  \multicolumn{1}{c}{Present} &%
  \multicolumn{1}{c}{Refs.\ \onlinecite{lyons_PRL2012,yan_APL2012}} &%
  \multicolumn{1}{c}{Present} &%
  \multicolumn{1}{c}{Refs.\ \onlinecite{lyons_PRL2012,yan_APL2012}} &%
  \multicolumn{1}{c}{Ref.\ \onlinecite{buckeridge_PRL2015}} \\
  \cline{1-2} \cline{3-4} \cline{5-7}
  \multirow{3}{*}{Mg$_\text{Ga}$}%
         & $E_\text{f}^0$         &  2.0       &  2.0    
                                  &  1.6       &  \multicolumn{1}{c}{--}      &  0.8 \\  
         & $E_\text{f}^{-1}$      &  2.4       &  2.3    
                                  &  2.0       &  \multicolumn{1}{c}{--}     &  2.2 \\  
         & $\varepsilon_{0/-1}$   &  0.38      &  0.26   
                                  &  0.38      &  0.26   &  1.40 \\ 
  \cline{1-2} \cline{3-4} \cline{5-7}
  \multirow{5}{*}{V$_\text{N}$}%
         & $E_\text{f}^{+3}$      & -1.7       & -1.0    %
                                  & -0.3       &  0.3    & -3.7 \\
         & $E_\text{f}^{+1}$      & -0.3       & -0.1    %
                                  &  1.1       &  1.2    & -0.1 \\
         & $E_\text{f}^{0}$       &  3.3       &  3.2    %
                                  &  4.7       &  4.3    & \multicolumn{1}{c}{--} \\
         & $\varepsilon_{+3/+1}$  &  0.70      &  0.47   %
                                  &  0.70      &  0.47   &  1.83 \\
         & $\varepsilon_{+1/0}$   &  3.63      &  3.25   %
                                  &  3.63      &  3.25   & \multicolumn{1}{c}{--} \\
  \cline{1-2} \cline{3-4} \cline{5-7}
  \multirow{3}{*}{Mg$_\text{Ga}$-V$_\text{N}$}%
         & $E_\text{f}^{+2}$      & -0.4       &  0.2    %
                                  &  0.5       &  1.2    & \multicolumn{1}{c}{--} \\
         & $E_\text{f}^{+2}$      &  1.2       &  2.0    %
                                  &  2.1       &  2.8    & \multicolumn{1}{c}{--} \\
         & $\varepsilon_{+2/0}$   &  0.80      &  0.87   %
                                  &  0.80      &  0.87   & \multicolumn{1}{c}{--} \\
  \cline{1-2} \cline{3-4} \cline{5-7}
  \multirow{1}{*}{Mg$_\text{inter}$}%
         & $E_\text{f}^{+2}$      &  1.1       & \multicolumn{1}{c}{--}      %
                                  & -0.2       & \multicolumn{1}{c}{--}      & \multicolumn{1}{c}{--} \\
\end{tabular}
\end{ruledtabular}
\end{table*}

We investigate the energetics of relevant defects acting as acceptors or donors 
during the growth process of GaN. As effective acceptor species, we only consider
the magnesium impurity substitutional to gallium. Among the possible counteracting
donors, we consider V$_\text{N}$, the complex Mg$_\text{Ga}$-V$_\text{N}$,
and the magnesium interstitial, Mg$_\text{inter}$. The relaxed structures of these
point defects are illustrated in Fig.\ \ref{fig:cell}. The defect formation energies
vs.\ Fermi energy are given Fig.\ \ref{fig:energies}.

In its neutral state, the structure of Mg$_\text{Ga}$ preserves the C$_{3v}$ 
symmetry of wurtzite [cf.\  Fig.\ref{fig:cell}(a)]. A hole is well localized on 
the axial N atom and its trapping is accompanied by a large polaronic lattice 
distortion, which results in a Mg$_\text{Ga}$-N bond stretched by 15\% compared
to a regular Ga-N bond. When an electron is added to the defective GaN:Mg$_\text{Ga}$ 
cell the hole is filled and all the Mg-N bonds become equivalent. The Mg$_\text{Ga}$
impurity gives rise to a 0/$-1$ acceptor level at 0.38 eV above the valence-band
maximum (VBM), in accord with reported experimental values lying in the range 
0.22--0.28 eV.\cite{madelung2004,monemar_PRL2009,monemar_PSSC2010,monemar_JAP2014}
We note that the hole localization is well described only when a hybrid 
density-functional approach is adopted.\cite{lany_APL2010,lyons_PRL2012,sun_PRB2014}
The use of semilocal density functionals yields a delocalized electronic state 
and a defect structure with four Mg-N bonds of similar length. 

The nitrogen vacancy (V$_\text{N}$) [cf.\ Fig.\ \ref{fig:cell}(b)] has long been 
considered the main counteracting donor defect. Indeed, among the native point 
defects, the nitrogen vacancy is the most stable one for a wide range of Fermi 
energies within the band-gap.\cite{neugebauer_PRB1994,boguslawski_PRB1995,limp_PRB2004,miceli_MEE2015} 
The nitrogen vacancy is found to be stable in the charge states $+1$ and $+3$, 
with a direct transition at 0.70 eV above the VBM. The neutral and $+2$ charge 
states are metastable. In the relaxed structures of the stable charge states, the
four nearest-neighbor Ga atoms are displaced away from the vacancy site. This 
effect becomes larger as the positive charge state of the vacancy increases.  
The difference between the formation energy of V$_\text{N}$ in N-rich and 
Ga-rich conditions corresponds to the formation energy of GaN. Our calculations 
give 1.4 eV, which favorably compares with the experimental enthalpy of formation
of 1.6 eV.\cite{ranade_JPCB2000} 

Another possible counteracting donor is the Mg$_\text{Ga}$-V$_\text{N}$
complex [cf.\ Fig.\ \ref{fig:cell}(c)]. The calculated formation energies show 
that this defect behaves like a double donor in deep $p$-type conditions and is 
neutral otherwise. The corresponding charge transition occurs at 0.80 eV above the 
VBM. The calculated formation energy implies a moderate defect concentration, in 
agreement with estimations based on positron annihilation 
spectroscopy.\cite{hautakangas_PRL2003,hautakangas_PRB2005}
These low densities rule out the Mg$_\text{Ga}$-V$_\text{N}$ complex as 
possible origin of the severe compensation observed in heavily doped samples 
(cf.\ Ref.\ \onlinecite{monemar_PSSC2010}).

We summarized in Table \ref{tab:energies} all the calculated formation energies
and charge transition levels. In particular, this table also contains results from
previous theoretical studies for comparison.\cite{lyons_PRL2012,yan_APL2012,buckeridge_PRL2015}
In general, our results are in very good agreement with those in 
Refs.\ \onlinecite{lyons_PRL2012,yan_APL2012}, but differ noticeably from those 
in  Ref.\ \onlinecite{buckeridge_PRL2015}. For instance, for Mg$_\text{Ga}$,
our defect level at 0.38 eV agrees well with the value of 0.26 eV found in 
Ref.\ \onlinecite{lyons_PRL2012}, but lies far away from the level of 1.404 eV 
reported in Ref.\ \onlinecite{buckeridge_PRL2015}. Similarly, for V$_\text{N}$ 
in N-rich conditions, we find formation energies of $1.1$ and $-0.3$ eV for the 
$+1$ and $+3$ charge states, respectively, in good agreement with the values of 
$\sim$1.2 and $\sim$0.25 eV from Ref.\ \onlinecite{yan_APL2012}, but in disaccord
with the values of $\sim$$-$0.15 and $\sim$$-$3.7 eV from 
Ref.\ \onlinecite{buckeridge_PRL2015}. We therefore do not confirm the low 
formation energies of V$_\text{N}$ found in the latter work. Furthermore, we 
remark that the present energies for V$_\text{N}$ obtained at the hybrid-functional
level also agree with those obtained at the semilocal level after proper alignment.\cite{miceli_MEE2015}

Our investigation also comprises the Mg positioned in an octahedral interstitial
site of GaN [Fig.\ \ref{fig:cell}(d)]. The interstitial Mg impurity is generally
discarded from the outset as an early theoretical study based on semilocal 
functionals found this defect at higher energies than the substitutional 
Mg$_\text{Ga}$.\cite{neugebauer_MRSP1995} 
We find that Mg$_\text{inter}$ behaves like a double donor irrespective of
the position of the Fermi level in the band gap (cf.\ Fig.\ \ref{fig:energies}),
and that it becomes noticeably more stable than the substitutional Mg$_\text{Ga}$
in $p$-type conditions. This is consequence of the downwards shift of the VBM 
achieved with hybrid functionals.\cite{AlkauskasPRL2008,alkauskas_PSSB2011,alkauskas_PRB2011} 
Our calculations therefore imply a stable interstitial state for the Mg impurity.
The Mg impurity in GaN is amphoteric and, depending on the Fermi level, it either
behaves like an acceptor in the substitutional site (Mg$_\text{Ga}$) or like a
double donor in the interstitial site (Mg$_\text{inter}$). The amphoteric nature
of the Mg impurity in GaN could lead to Fermi level pinning.%
\cite{walukiewicz_JVSTB1987,*walukiewicz_PRB1988,colleoni_JPCM2014}

In case the Mg interstitial easily diffuses out of the system, its role as a
compensating donor would not apply. This could be particularly critical at
the growth temperature of 1300 K. An experimental study has determined 
the diffusion coefficient of Mg in GaN to follow an Arrhenius behavior, 
characterized by an acrtivation energy of 1.9 eV and a prefactor of 
$D_0= 2.8\times 10^{-7}$ cm$^2$/s.\cite{benzarti_JCG2008} On this basis, 
we estimate a linear diffusion length of 0.13 $\mu$m for a period of two hours, 
corresponding to typical growth conditions. This distance is short compared 
to typical layer thicknesses of 1 to 2 $\mu$m. Hence, the Mg interstitial
is expected to remain trapped within the sample.

\begin{figure}
\centering
\includegraphics[width=8cm]{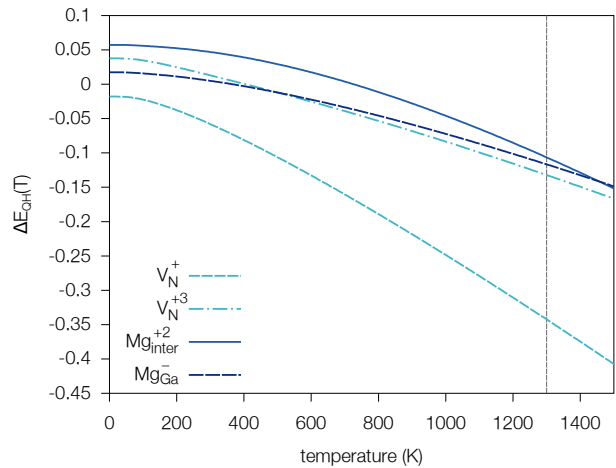}
\caption{Calculated correction $\Delta E_\text{QH}(T)$ to the formation-energy 
         formulation at zero temperature for V$_\text{N}^{+}$, 
         V$_\text{N}^{+3}$, Mg$_\text{inter}^{+2}$, and 
         Mg$_\text{Ga}^{-}$. The vertical line indicates 
         the growth temperature of 1300 K. }
\label{fig:free-energies}
\end{figure}

To determine the corrections to the formation-energy formulation at zero
temperature, we calculate $\Delta E_\text{QH}(T)$ given in Eq.\ (\ref{eq:qh}) for 
Mg$_\text{Ga}$, Mg$_\text{inter}$, and V$_\text{N}$ in their stable charge states.
For each defect, we evaluate the $\Gamma$-point phonons of the supercell using 
linear-response theory as implemented in {\sc Q}uantum-{\sc Espresso}.\cite{baroni_RMP2001}
In Fig.\ \ref{fig:free-energies}, the calculated contributions, $\Delta E_\text{QH}(T)$,
as a function of temperature are given. The displayed contribution includes both the 
zero point motion and the temperature-dependent term [cf.\ Eqs.\ (\ref{eq:qhapprox})
and (\ref{eq:qh})]. As one can notice, at 1300 K, corresponding to the typical 
growth temperature used in MOVPE, all the defects under investigation undergo an
entropic stabilization. For the nitrogen vacancy in its charge state $+1$, we 
observe a significant energy gain of 0.3 eV. For the other defects, the stabilization
is smaller and amounts to $\sim$0.1 eV.

\section{Self-compensation mechanisms} \label{sec:self}

The energetics in Fig.\ \ref{fig:energies} suggest that both V$_\text{N}$ and 
Mg$_\text{inter}$ could counteract the $p$-doping process. The free energies 
of formation $F^\text{f}$ given in Eq.\ (\ref{eq:freeener1}) cannot directly be compared
with experimental data, but determine the defect concentrations at thermodynamic
equilibrium. For a given acceptor (A) or donor (D) impurity, the equilibrium 
concentration is obtained by minimizing the total configurational free energy of
the system and reads
\begin{equation}
N_\text{A/D} = N_\text{S} \exp{[ - F^\text{f}_\text{A/D}(\mu_\text{F})/k_\text{B}T] },
\label{eq:ADconc}
\end{equation}
where $N_\text{S}$ is the number of sites per volume in which the point defect
could occur and $F^\text{f}_\text{A/D}(\mu_\text{F})$ is the free energy 
of formation which depends on the Fermi energy $\mu_\text{F}$ in the case of 
charged defects. In the following, we consider self-compensation models which include
Mg$_\text{Ga}$ as an acceptor and both Mg$_\text{inter}$ and V$_\text{N}$ 
as compensating donors. For a given Mg doping density and growth temperature, the
respective Boltzmann factors then provide us with the V$_\text{N}$ concentration
and the relative abundances of Mg$_\text{Ga}$ and Mg$_\text{inter}$. Since 
the formation free energies of charged defects depend on the Fermi level,
the defect concentrations need to be determined self-consistently along 
with the Fermi level and the hole concentration $p$. Thus, in order to evaluate
the compensating role of Mg$_\text{inter}$ and V$_\text{N}$ upon Mg doping, 
we resort to the equations of semiconductors dominated by impurities, which allow 
us to directly link our hybrid-functional results with experimental observations.
For a non-degenerate semiconductor at thermodynamic equilibrium, we have:\cite{blakemore1962}
\begin{align}
& N_\text{A} = N_\text{A}(\mu_\text{F})  \quad ; \qquad  N_\text{D} = N_\text{D}(\mu_\text{F})     \label{eq:acceptordonor} \\
& p = - \frac{(N_\text{D} + \mathcal{K})}{2} + %
        \sqrt{ \frac{(N_\text{D} + \mathcal{K})^2}{4} + \mathcal{K} (N_\text{A} - N_\text{D}) } \label{eq:hole} \\
& \mu_F = E_\text{v} - k_\text{B} T \ln \left(\frac{p}{\mathcal{N}_\text{v}} \right), \label{eq:fermi}
\end{align}
where $ \mathcal{K} = (\mathcal{N}_\text{v}/\beta) \, e^{-E_\text{A}/k_\text{B}T}$, 
$ \mathcal{N}_\text{v} = 2 \left({2\pi m_\text{v} k_\text{B} T}/{h^2} \right)^{3/2}$
is the effective density of states in the valence band, and $E_\text{A}$ the 
activation energy of the acceptor state. In these expressions, the valence band 
degeneracy factor $\beta=4$ (Ref.\ \onlinecite{look1989}) and the valence-band 
effective mass $m_\text{v}=0.8$ (Ref.\ \onlinecite{rinke_PRB2008}) are kept 
fixed. In view of possible HSE-related inaccuracies,\cite{chen_JPCM2015} we 
rigidly shift the calculated band edges to match the activation energy of 0.16 eV
used in the analysis of the experimental data of Ref.\ \onlinecite{kaufmann_PRB2000} 
that we aim at interpreting. Moreover, we neglect any effect resulting from the 
dependence of $E_\text{A}$ on doping concentration.\cite{brochen_APL2013}

In all the presented growth models, we assume Ga-rich conditions which are the 
experimental thermodynamic conditions at which $p$-doped GaN is commonly grown.
Within this assumption the nitrogen vacancy concentration is determined 
self-consistently based on the energetics shown in Fig.\ \ref{fig:energies}(a) 
including free-energy corrections at the given growth temperature, as shown in 
Fig.\ \ref{fig:free-energies}. Both stable charge states of V$_\text{N}$, $+1$ 
and $+3$, are taken into account to solve the charge neutrality equations.

In this work the relevant physical quantities are presented as a function of the
Mg-doping concentration. Unlike gallium and nitrogen species, the magnesium 
chemical potential is therefore assumed to vary. The variation of $\mu_\text{Mg}$ 
does not influence the determination of the relative abundances of Mg$_\text{Ga}$
and Mg$_\text{inter}$. In fact, given a total Mg-doping concentration, the 
fraction of Mg going into interstitial or substitutional sites can be expressed 
as the ratio of the respective defect Boltzmann factors, which does not depend 
on the magnesium chemical potential.

\subsection{Equilibrium conditions}
\label{subsec:equilibrium}

\begin{figure}
\centering
\includegraphics[width=8cm]{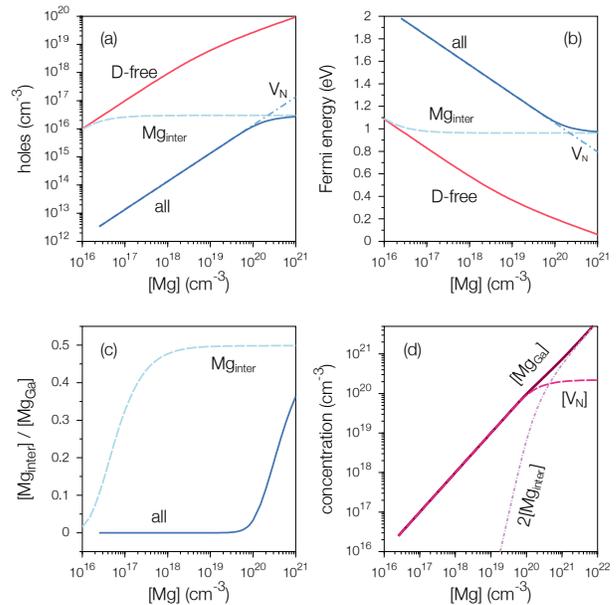}
\caption{(a) Hole density and (b) Fermi level as a function of the Mg doping 
         concentration obtained at thermodynamic equilibrium for a growth 
         temperature of 1300 K. The different curves refer to a donor-free 
         model (D-free) and to three compensated models, with Mg$_\text{inter}$
         donors only, with V$_\text{N}$ donors only or with the combined 
         effect of the Mg$_\text{inter}$ and V$_\text{N}$ donors (all).
         In panel (c), the ratio [Mg$_\text{inter}$]/[Mg$_\text{Ga}$]
         is shown as a function of the doping concentration. Panel (d) shows
         the defect concentrations for the case in which both Mg$_\text{inter}$
         and V$_\text{N}$ act as compensating donors (all).}
\label{fig:equilibrium}
\end{figure}

\begin{figure}
\centering
\includegraphics[width=8cm]{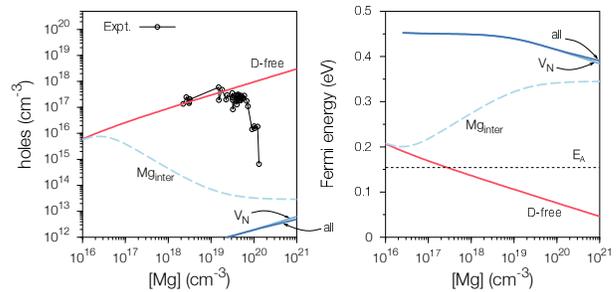}
\caption{(a) Hole density and (b) Fermi energy calculated at 300 K while
         preserving the acceptor and donor concentrations obtained at 1300 K.
         Experimental data from Ref.\ \onlinecite{kaufmann_PRB2000} are reported 
         in (a) for comparison.}
\label{fig:hF-equilibrium}
\end{figure}

The solution of Eqs.\ (\ref{eq:acceptordonor})--(\ref{eq:fermi}) gives the 
acceptor, the donor, and the hole concentrations along with the Fermi energy as 
a function of the Mg doping concentration at thermodynamic equilibrium. For a 
typical growth temperature of 1300 K,\cite{kaufmann_PRB2000} we obtain the results
given in Fig.\ \ref{fig:equilibrium}. In particular, in Fig.\ \ref{fig:equilibrium},
we distinguish four different compensating models: the ideal case, in which the 
system is donor-free and all the Mg atoms go into substitutional Ga sites, and 
three self-compensating models with V$_\text{N}$ donors only, with Mg$_\text{inter}$
donors only, or with the combined effect of both V$_\text{N}$ and Mg$_\text{inter}$
donors. In the donor-free case, the absence of donors ensures that the hole 
concentration reaches its highest value for any Mg doping density and sets an 
upper limit for more realistic cases. When the formation of both V$_\text{N}$ and
Mg$_\text{inter}$ compensating donors is allowed, the achieved hole densities are
radically lower. In particular, we note that for a low Mg doping density the 
nitrogen vacancies are the dominant counteracting donors. As compared to the 
donor-free case, their presence keeps the Fermi level at higher values in the 
band gap for a given Mg doping concentration [Fig.\ \ref{fig:equilibrium} (b)]. 
However, as soon as the Fermi energy comes closer to the energy level where the 
formation energies of Mg$_\text{Ga}$ and Mg$_\text{inter}$ become equal [the 
plateau at $\sim$1 eV in Fig.\ \ref{fig:equilibrium}(b)], i.e.\ for Mg doping 
concentrations around $10^{20}$ cm$^{-3}$, the Mg$_\text{inter}$ interstitials 
proliferate giving rise to self-compensation. When only Mg$_\text{inter}$ are 
considered as compensating donors, the same proliferation of Mg$_\text{inter}$ 
is observed, but at Mg concentrations as small as $\approx 10^{17}$ cm$^{-3}$.
Upon the abrupt proliferation of Mg interstitials, the Fermi level pins and the 
ratio [Mg$_\text{inter}$]/[Mg$_\text{Ga}^-$] assumes the value of 1/2 
[cf.\ Figs.\ \ref{fig:equilibrium} (b) and (c)]. For a further increase of the 
Mg doping level, both the hole concentration [Figs.\ \ref{fig:equilibrium}(a)] 
and the Fermi energy [Figs.\ \ref{fig:equilibrium}(b)] remain constant and the 
doping process is abruptly arrested. In this regime, the Mg$_\text{Ga}$ and 
Mg$_\text{inter}$ concentrations increase with a constant ratio 2:1. At variance,
the vacancy density remains constant as determined by the pinned Fermi level. We
note that in this phenomenology the amphoteric nature of the Mg impurity is critical.
Indeed, would V$_\text{N}$ be the only compensating donor, the doping process 
would have continued, reaching higher hole concentration and lower Fermi levels 
with increasing Mg doping levels, as shown in Figs.\ \ref{fig:equilibrium}(a) and
(b), respectively.

In Fig.\ \ref{fig:equilibrium}(d), we plot the defect concentrations as a function
of the Mg doping density for the model in which both V$_\text{N}$ and 
Mg$_\text{inter}$ act as compensating donors. We notice that at low Mg doping 
densities, the V$_\text{N}$ donor concentration closely follows the increase in 
Mg$_\text{Ga}$ concentration, resulting in an immediate compensation. This trend
is reflected by the behavior of the Fermi energy, which decreases slowly with 
increasing Mg doping density [cf.\ Fig.\ \ref{fig:equilibrium}(b)]. At variance,
the Mg$_\text{inter}$ concentration undergoes a sudden increase leading to the 
abrupt pinning of the Fermi level. The explanation of such a behavior rests on 
the faster rate of increase of [Mg$_\text{inter}$] with respect to [V$_\text{N}$].
Indeed, while [V$_\text{N}$] changes as described by Eq.\ (\ref{eq:ADconc}), 
[Mg$_\text{inter}$] varies rapidly when the difference in formation energy between
Mg$_\text{inter}$ and Mg$_\text{Ga}$ becomes comparable to $k_\text{B}T$. More 
specifically, through the use of the Boltzmann factors, one obtains  
\begin{equation}
[\text{Mg}_\text{inter}] =  \frac{ [\text{Mg}] } %
{ 1 + \exp \left( \tfrac{ 3 ( \mu_\text{F} - \mu_\text{F}^* ) }{k_\text{B}T} \right) },
\label{eq:concinter}
\end{equation}
where $\mu_\text{F}^*$ is the Fermi level position at which the formation free 
energies of Mg$_\text{inter}$ and Mg$_\text{Ga}$ are equal.

For a direct comparison between theory and experiments, we calculate the equilibrium
hole density at room temperature assuming acceptor and donor concentrations as 
obtained at the growth temperature of 1300 K. This procedure attempts to capture
the effects of the rapid thermal quench undergone by the samples, upon which they
would preserve the equilibrium defect concentrations achieved at the growth 
temperature. In Fig.\ \ref{fig:hF-equilibrium}, we give the hole concentrations and 
the Fermi level at room temperature, as determined through Eqs.\ (\ref{eq:hole})
and\ (\ref{eq:fermi}) within the various self-compensation models. From 
Fig.\ \ref{fig:hF-equilibrium}(a), one notices that, even for a donor-free case,
the hole concentration is much lower than the total [Mg] concentration. This 
effect results from the reduction of the ionized acceptors [Mg$_\text{Ga}^{-}$],
when the decreasing $\mu_\text{F}$ reaches the ionization energy $E_A$. Indeed, 
at a given temperature, [Mg$_\text{Ga}^-$] can be expressed as
\begin{equation}
\label{eq:mggaionized}
[\text{Mg}^-_\text{Ga}] = \frac{[\text{Mg}_\text{Ga}]}{1+ \beta \exp \left( - \frac{\mu_\text{F} - E_A}{k_\text{B}T} \right)},
\end{equation}
where [Mg$_\text{Ga}$]=[Mg$_\text{Ga}^0$]+[Mg$_\text{Ga}^-$] is the total 
concentration of magnesium substitutional to gallium. The high value of $E_A$, 
as compared to the thermal energy at room temperature, requires heavy doping to 
achieve high hole concentrations. 

For comparison, we also report in Fig.\ \ref{fig:hF-equilibrium}(a) the experimental 
hole densities measured in Ref.\ \onlinecite{kaufmann_PRB2000}. In the most 
realistic case, when both Mg$_\text{inter}$ and V$_\text{N}$ are acting as 
compensating donors, the calculated hole density differs from the experimental 
values by several orders of magnitude. This implies that a model based on the 
achievement of bulk equilibrium properties does not apply. In fact, the 
experimental data show that as long as the Mg-doping concentrations remain below
a threshold value of about $10^{19}$ cm$^{-3}$, the achieved hole densities agree
with those pertaining to the donor-free model [cf. Fig.\ \ref{fig:hF-equilibrium}(a)].
Beyond the threshold concentration of $\sim$$10^{19}$ cm$^{-3}$, the experimental
data indicate that the compensation effects intervene severely and suddenly.%
\cite{kaufmann_PRB2000,namkoong_APL2008,brochen_APL2013,malinverni_APL2014}
The sudden nature of this behavior contrasts with the gradual way in which 
V$_\text{N}$ compensates the hole concentration at equilibrium conditions 
[cf.\ Figs.\ \ref{fig:equilibrium}(a) and (d)]. At variance, the amphoteric nature
of the Mg impurity and the achievement of Fermi-level pinning through a sudden 
proliferation of Mg interstitials appear more appropriate to the experimental 
phenomenology [cf.\ Fig.\ \ref{fig:equilibrium} (d)].

\subsection{Non-equilibrium models}
\label{subsec:non-equilibrium}

\begin{figure}
\centering
\includegraphics[width=8cm]{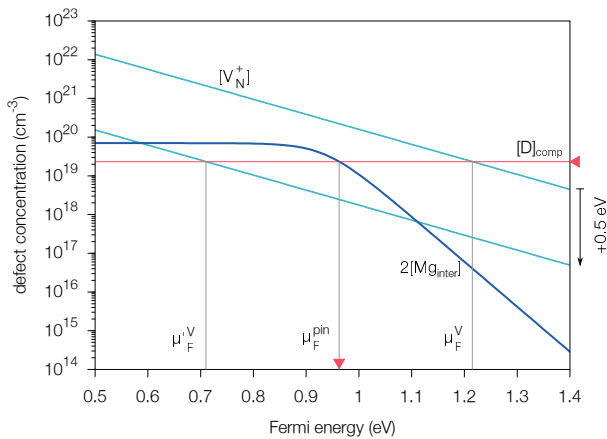}
\caption{Concentration of nitrogen vacancy and Mg interstitial as a function of
         Fermi level. [Mg$_\text{inter}$] is obtained through Eq.\ (\ref{eq:concinter}) 
         at the [Mg]$_\text{th}$ doping threshold 
         ($\approx 3.5 \times 10^{19}$ cm$^{-3}$, from 
         Ref.\ \onlinecite{kaufmann_PRB2000}). The factor of 2 accounts for the
         multiplicity of the Mg$_\text{inter}$ donor. We also show the concentration
         of the nitrogen vacancy when it is destabilized by 0.5 eV. The horizontal
         line represents the compensating donor concentration $[\text{D}]_\text{comp}$.}
\label{fig:donors}
\end{figure}

In this section, we attempt to identify which donor defect is the dominating compensating
defect at the origin of the drop-off in the hole concentration upon Mg doping in GaN,
observed at a typical threshold concentration given by $[\text{Mg}]_\text{th}\approx 
3.5\times 10^{19}$ cm$^{-3}$ (Ref.\ \onlinecite{kaufmann_PRB2000}).
We consider either Mg$_\text{inter}$ or V$_\text{N}$ defects as counteracting donors.
More specifically, we assume that Mg incorporates either interstitially or substitutionally:
\begin{equation}\label{eq:Mg}
[\text{Mg}]= [\text{Mg}_\text{inter}] + [\text{Mg}_\text{Ga}]. 
\end{equation}
In our description, Mg$_\text{inter}$ is always fully ionized,
while V$_\text{N}$ only occurs in the charge states +1 for the  
Fermi energies under consideration (cf.\ Fig.\ \ref{fig:energies}).
The charge compensation equation then gives:
\begin{equation}\label{eq:compensation}
[\text{Mg}_\text{Ga}^{-}] = 2[\text{Mg}_\text{inter}] + [\text{V}_\text{N}] +  p,
\end{equation}
where $p$ is the hole density which vanishes at the threshold. Therefore,
one can distinguish two different regimes in which either the nitrogen vacancy or 
the Mg interstitial clearly dominate the counteracting action: 
\begin{align}
&2[\text{Mg}_\text{inter}] \ll  [\text{V}_\text{N}]\quad\Rightarrow\quad 
[\text{Mg}_\text{Ga}^{-}]=[\text{V}_\text{N}],  \label{eq:1v}\\
&2[\text{Mg}_\text{inter}] \gg  [\text{V}_\text{N}]\quad\Rightarrow\quad
[\text{Mg}_\text{Ga}^{-}]=2[\text{Mg}_\text{inter}]. \label{eq:2v}
\end{align}

In order to understand the origin of the observed drop-off, we aim at identifying 
the compensating donor concentration $D_\text{comp}$, beyond which the efficiency of the $p$ 
doping process drastically decreases. According to Eqs.\ (\ref{eq:1v}) and (\ref{eq:2v}),
$D_\text{comp}$ corresponds to $[\text{Mg}_\text{Ga}^{-}]$ at 300 K when the Mg doping 
density reaches the threshold concentration $[\text{Mg}]_\text{th}$, and does not depend 
on the specific counteracting donor.  For convenience, we estimate $[\text{D}]_\text{comp}$  
in case the dominating donor is the Mg interstitial. In this case, 
$[\text{D}]_\text{comp}$ is equal to $2[\text{Mg}_\text{inter}]$ [cf.\ Eq.\ (\ref{eq:2v})] and  
can be evaluated indifferently at 300 K or at 1300 K, since all interstitial Mg 
are always fully ionized. 
At the growth temperature of 1300 K, the defect concentrations are set by equilibrium conditions. 
The Mg interstitials occur when the Fermi energy reaches the pinning level. 
Therefore, we assume that the pinning occurs in correspondence of the
experimental doping threshold $[\text{Mg}]_\text{th}$:
\begin{equation}
\mu_\text{F} ([\text{Mg}]_\text{th}) = \mu_\text{F}^\text{pin}
\end{equation}
In correspondence of the pinning, the hole density becomes negligible with respect to
the $2[\text{Mg}_\text{inter}]$ and can thus be neglected in Eq.\ (\ref{eq:compensation}),
giving:
\begin{equation} \label{eq:ratio2:1}
[\text{Mg}_\text{Ga}^{-}]_\text{1300~K} = 2[\text{Mg}_\text{inter}] \approx [\text{Mg}_\text{Ga}],
\end{equation}
where the second approximate equality holds because all the substitutional Mg are
ionized at 1300 K, i.e.\ $[\text{Mg}_\textrm{Ga}^0]\approx 0$, since $\mu_\text{F} \gg 
E_\text{A}$ [cf.\ Eq.\ (\ref{eq:mggaionized})].
Combining Eqs.\ (\ref{eq:ratio2:1}) and (\ref{eq:Mg}), we then find 
the following expression for $[\text{D}]_\text{comp}$: 
\begin{equation}
[\text{D}]_\text{comp} = \frac{2}{3} [\text{Mg}]_\text{th}.
\label{eq:Dcomp}
\end{equation}

The previous analysis is graphically illustrated in Fig.\ \ref{fig:donors}. Here, the 
defect concentrations are plotted as a function of Fermi energy for a temperature of 1300 K. 
The Mg$_\text{inter}$ concentration is calculated through Eq.\ (\ref{eq:concinter}) 
at the threshold [Mg]$_\text{th}$ doping density,\cite{kaufmann_PRB2000} while the V$_\text{N}$ 
concentration is determined through Eq.\ (\ref{eq:ADconc}) for Ga-rich conditions. 
The horizontal line represents the compensating donor concentration $[\text{D}]_\text{comp}$.
From Fig.\ \ref{fig:donors}, we infer that the Fermi level $\mu_\text{F}^\text{V}$ at which 
[V$_\text{N}$] reaches $[\text{D}]_\text{comp}$ is larger than the Fermi level 
$\mu_\text{F}^\text{pin}$ at which [Mg$_\text{inter}$] reaches $[\text{D}]_\text{comp}$.  
Since the Fermi energy decreases during growth, the condition at $\mu_\text{F}^\text{V}$ 
realizes before that at $\mu_\text{F}^\text{pin}$, implying that the nitrogen vacancy 
is the dominating compensating donor. However, we note that in the hypothetical case
in which the nitrogen vacancy is destabilized by 0.5 eV, the situation would be 
reversed and the Mg$_\text{inter}$ would be the principal counteracting donor.

In Sec.\ \ref{subsec:equilibrium}, we saw that the equilibrium conditions at growth 
temperature cannot explain the sudden drop-off in the hole density,
experimentally observed at $[\text{Mg}]_\text{th}\approx 3.5\times 10^{19}$ cm$^{-3}$. 
The experimental behavior could be reconciled with a Fermi level located at higher
energies in the band gap. However, bulk equilibrium conditions would draw the Fermi
energy to lower values as the result of self-consistency, leading to more favorable
conditions for donor generation and charge compensation. Since the growth takes
place at the surface, we abandon the principle that the Fermi level position is
determined by sole bulk conditions and assume that it could be affected by specific
conditions occurring at the surface. Indeed, due to impurity incorporation, a downwards 
band-bending has been observed at $p$-type GaN surfaces.\cite{barbet_APL2008,sezen_APL2011}
For instance, the measured band bending reaches the value of $-$1.58 eV 
for a Mg doping density of $\sim$$5\times 10^{17}$ cm$^{-3}$ (Ref.\ 
\onlinecite{barbet_APL2008}). Such an effect would lead to Fermi levels located at 
higher energy than those resulting from bulk equilibrium conditions 
[cf.\ Fig.\ \ref{fig:equilibrium}(b)]. The band bending extends over a surface layer of 
several hundreds angstroms, in which we propose the defect incorporation takes place. 
We assume that the formation energies in this region do not differ from their bulk 
value, neglecting thereby variations that might occur directly at the surface.
In the following, we adopt such non-equilibrium models to interpret the experimental
evidence. 

In the next two subsections, we separately discuss the cases in which the nitrogen 
vacancy and the Mg$_\text{inter}$ are the dominating donors. 
In particular, we impose that our models reproduce the experimental drop-off 
in the hole density occurring at $[\text{Mg}]_\text{th}$. From such a description,
we then infer the dependence of the Fermi level vs.\ the Mg doping density. We also 
obtain the defect concentrations of the relevant donors from Eqs.\ (\ref{eq:acceptordonor}).

\subsubsection{Nitrogen vacancy as dominating donor}
\label{subsub:vac}

\begin{figure}
\centering
\includegraphics[width=8cm]{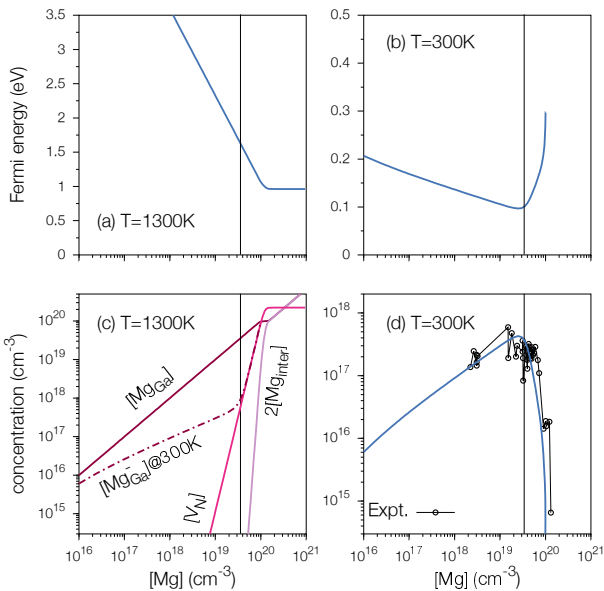}
\caption{Non-equilibrium vacancy-dominated self-compensation model designed to 
         interpret the experimental density of holes vs Mg doping concentration 
         [from Ref.\ \onlinecite{kaufmann_PRB2000}, panel (d)]: 
         Fermi level (a) at 1300 K and (b) at 300 K, (c) defect concentrations 
         at 1300 K, and (d) hole density at 300 K as a function of
         Mg doping concentration. }
\label{fig:vmodel}
\end{figure}

Following the outcome of our calculations (cf.\ Fig.\ \ref{fig:donors}), we first
consider a self-compensation model in which the V$_\text{N}$ are the dominating 
compensating donors. In Fig.\ \ref{fig:vmodel}, we show the evolution of various
physical quantities when the model is designed to reproduce the experimental 
behavior of the hole density vs.\  Mg doping concentration [cf.\ Fig.\ \ref{fig:vmodel}(d)]. 
Figure \ref{fig:vmodel}(a) gives the required dependence of the Fermi level during
the growth at 1300 K, when the defect structures are formed 
[cf.\ Fig.\ \ref{fig:vmodel}(c)]. The hole density at room temperature as adopted 
in our model is compared with the experimental one in Fig.\ \ref{fig:vmodel}(d).

As shown in Fig.\ \ref{fig:vmodel}(a), the Fermi level at 1300 K monotonically 
decreases with Mg doping concentration in a similar way as under equilibrium 
conditions [cf.\ Fig.\ \ref{fig:equilibrium}(b)], but generally remains at higher
energies until it reaches the pinning level. In turn, the Fermi energy
determines through Eq.\ (\ref{eq:ADconc}) the rate at which the compensating 
Mg$_\text{inter}$ and V$_\text{N}$ are formed, as displayed in Fig.\ \ref{fig:vmodel}(c). 
In Figs.\ \ref{fig:vmodel}(b) and (d), we show the Fermi level and the hole
density at 300 K, respectively, while keeping the concentrations of the defect structures
achieved at the growth temperature. At 300 K, all Mg$_\text{inter}$ and V$_\text{N}$ 
remain ionized, while the fraction of activated Mg$_\text{Ga}$ decreases in
a significant way due to the lower temperature [cf.\ Eq.\ (\ref{eq:mggaionized})]. 
As the Fermi level decreases with Mg doping density, the V$_\text{N}$ concentration
increases until it becomes comparable to the Mg$_\text{Ga}^{-}$ concentration, 
when a drastic decrease of the hole density is observed. 
In the present model, the nitrogen vacancy is thus the dominating donor defect 
leading to severe self-compensation. As shown in Fig.\ \ref{fig:vmodel}(c),
the proliferation of vacancies strongly counteracts the doping action played by 
the ionized Mg$_\text{Ga}^-$.

By construction, the present non-equilibrium model reproduces the experimental
results [see Fig.\ \ref{fig:vmodel}(d)]. Its validity should therefore be 
assessed by critically analyzing the behavior of the relevant physical quantities.
In particular, we notice that the present model implies a sudden and abrupt 
proliferation of nitrogen vacancies, in sharp contrast with the behavior 
observed in equilibrium conditions for [V$_\text{N}$] (compare
Fig.\ \ref{fig:vmodel} and Fig.\ \ref{fig:equilibrium}). 

\subsubsection{Magnesium interstitial as dominating donor}
\label{subsub:mg}

\begin{figure}
\centering
\includegraphics[width=8cm]{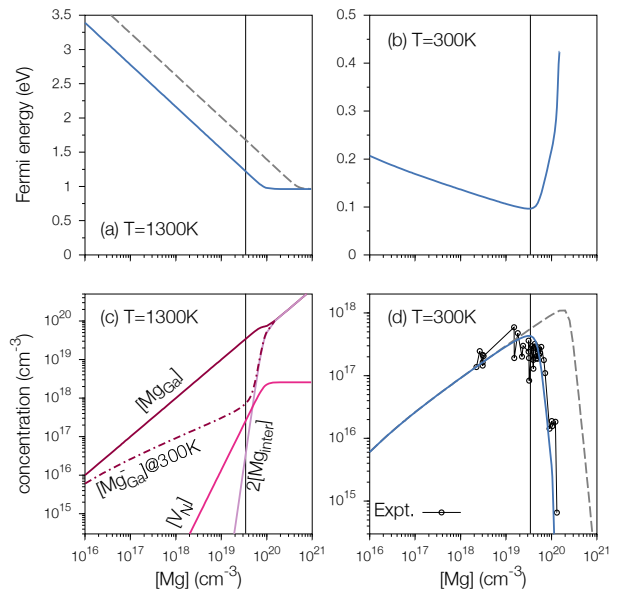}
\caption{Non-equilibrium Mg$_\text{inter}$-dominated self-compensation model designed to
         interpret the experimental density of holes vs Mg doping concentration
         [from Ref.\ \onlinecite{kaufmann_PRB2000}, panel (d)]:
         Fermi level (a) at 1300 K and (b) at 300 K, (c) defect concentrations
         at 1300 K, and (d) hole density at 300 K as a function of
         Mg doping concentration. In panels (a) and (d), the dashed curves 
         indicate a case, in which the drop-off density threshold is shifted 
         to a higher value.}
\label{fig:imodel}
\end{figure}

In this subsection, we consider a non-equilibrium model in which the dominating 
donor defects are the Mg$_\text{inter}$. As seen in Fig.\ \ref{fig:donors}, our 
calculations indicate that this can be achieved by destabilizing the vacancy by 
0.5 eV. This condition can be realized intentionally in experimental setups through
the use of high nitrogen partial pressures during growth.%
\cite{ambacher_JVSTB1996,ambacher_JJAP1998,ganchenkova_PRL2006}
This condition could also occur unintentionally in case the thermodynamic conditions
do not specifically correspond to the extreme Ga-rich conditions (cf.\ Figs.\ \ref{fig:therm}
and \ref{fig:energies}). Finally, typical density-functional-theory errors generally
amount to a few tenths of electronvolt, but larger errors cannot be ruled out. 
Therefore, the occurrence of this scenario should be taken under consideration.  

From the analysis at the beginning of Sec.\ \ref{subsec:non-equilibrium}, we expect that
a destabilization of the vacancy by 0.5 eV should cause the Mg interstitial
to become the compensating donor defect, cf.\ Fig.\ \ref{fig:donors}. 
Following the same procedure as for the case in which the nitrogen vacancy is the dominant 
compensating donor, we ensure that the model reproduces the drop-off in the hole density
as observed experimentally [Fig.\ \ref{fig:imodel}(d)] and monitor the behavior of the 
relevant physical quantities, such as the defect concentrations and the Fermi level. 

The evolution of the Fermi energy governing the growth process 
at 1300 K is shown in Fig.\ \ref{fig:imodel}(a). The Fermi level at 1300 K decreases 
in a smoother way with respect to the vacancy-dominated model, showing a closer 
resemblance with the behavior observed under equilibrium conditions 
[cf.\ Fig.\ \ref{fig:equilibrium}(a)]. The decrease stops when  
the [Mg$_\text{inter}$]/[Mg$_\text{Ga}^-$] ratio reaches the value of 1/2, 
corresponding to a sudden proliferation of Mg$_\text{inter}$.
Upon the pinning of the Fermi level, the concentration of V$_\text{N}$
reaches a plateau, as can be seen in Fig.\ \ref{fig:imodel}(c).
At the threshold Mg doping density, the hole drop-off is the result of
the compensation of the ionized acceptor concentration [Mg$_\text{Ga}^-$] at 300 K 
by the Mg interstitial concentration 2[Mg$_\text{inter}$] 
[Fig.\ \ref{fig:imodel}(c)]. In Figs.\ \ref{fig:imodel}(b) and (d), we show the 
associated Fermi level and hole density as found at 300 K.  

As remarked for the vacancy-dominated model, the validity of the model should be
assessed through the behavior of the relevant physical quantities. 
Unlike for the vacancy-dominated model, the sudden proliferation of 
interstitials can be explained in a natural way. Indeed, also 
in equilibrium conditions, the concentration of interstitials 
undergoes a sudden increase as a consequence of Fermi level pinning, which 
occurs when the formation energy of Mg$_\text{inter}$ and Mg$_\text{Ga}$ are 
approximately equal. The abundances of Mg$_\text{inter}$ and Mg$_\text{Ga}$
then only depend on their relative energy. At variance, in the vacancy-dominated model,
the abundance of nitrogen vacancies in equilibrium conditions depends on the 
V$_\text{N}$ formation energy which does not undergo abrupt variations 
as a function of Fermi level. We also remark that the decay of the
Fermi level at 1300 K in the present interstitial-dominated model
resembles more closely the behavior observed in equilibrium conditions
than in the vacancy-dominated model. In view of these considerations,
the abrupt rise of the compensating donor concentration in 
correspondence of the Mg doping threshold appears more compatible
with a response due to Mg interstitials than to nitrogen vacancies.

In the case of a vacancy-dominated self-compensation mechanism, the 
destabilization of the vacancy might improve the doping efficiency.
At variance, in a self-compensation mechanism dominated by Mg interstitials, 
a change of the thermodynamic conditions would only lead to a small shift 
of the pinned Fermi level without affecting the overall mechanism. 

\section{Conclusions}
\label{sec:conclusions}
Using a hybrid functional approach, we addressed the energetics of point 
defects and impurities in GaN that could play an important role in the 
self-compensation process occurring upon high levels of Mg doping. 
In particular, our calculations account for the free-energy of
formation of the nitrogen vacancy and of several Mg-related defects.
Our calculations revealed that the Mg impurity in GaN is amphoteric.
It behaves as a single acceptor when substitutional to Ga and as a double
donor when it occupies an interstitial site. Our study suggests that only the 
Mg interstitial and the nitrogen vacancy could act as compensating donors
upon Mg doping.

Using the calculated free energies of formation, we then used the equations 
of semiconductors dominated by impurities to establish a link with experimental
observations. These equations were solved self-consistently at thermodynamic 
equilibrium resulting in the determination of the donor and the acceptor 
concentrations, the Fermi level position, and the hole density as a function 
of the Mg doping concentration. Our results indicate that the defect concentrations
found under equilibrium conditions are unable to account for the drop-off in 
the hole density observed experimentally. 

We then studied non-equilibrium models which account for the drop-off in the 
hole density by construction and analyzed the behavior of the relevant 
physical properties, such as defect concentrations and the Fermi level.
In particular, we considered two scenarios in which either the nitrogen 
vacancies or the magnesium interstitials act as the dominant compensating donors. 
In both cases, the drop-off in the hole density could only be explained 
by a sudden proliferation of donor defects. In the case of the vacancy-dominated
mechanism, this sudden proliferation contrasts with the behavior found in equilibrium 
conditions and lacks a physical interpretation. At variance in the case of
the interstitial-dominated mechanism, the sudden proliferation is similar to
the one observed in equilibrium conditions and stems from the occurrence
of Fermi level pinning. 

These considerations favor the interpretation in which the dominant compensating 
donors are Mg interstitials. As long as the Fermi level is high in the band
gap, the Mg dopants enter the sample as substitutional impurities.  
Their $p$-type doping action then moves the Fermi level towards lower values.
When the formation energies of interstitial and substitutional Mg become 
approximately equal, the concentration of Mg interstitials suddenly rises and
the Fermi level is pinned through a feedback mechanism.
Hence, in this scenario, the amphoteric nature of the Mg impurity is critical 
to explain the drop-off in the hole density observed experimentally. 

Unlike the vacancy-dominated mechanism, in which variations of thermodynamic growth 
conditions could drastically impact the occurrence of compensation, the 
interstitial-dominated mechanism remains fairly insensitive to such variations
leading to at most a small shift of the pinned Fermi level. 
However, the Mg$_\text{inter}$-driven self-compensation discussed in this work 
leaves open the possibility of achieving $p$-doped GaN samples with higher hole 
concentrations. For this purpose, it is necessary to extrinsically 
control the Fermi energy during growth in such a way that the pinning of the Fermi
level is reached at a higher level of Mg doping. This could for instance be achieved
by increasing the electron density via UV illumination\cite{bryan_JEM2013} or 
by electron beam irradiation. 
We illustrate the effect of such interventions by rigidly shifting
the evolution of the Fermi energy as shown in Fig.\ \ref{fig:imodel}(a).
By consequence, the compensation due to Mg$_\textrm{inter}$ would activate at higher Mg
doping density and the hole density at room temperature could grow to higher values 
before reaching the drop-off [Fig.\ \ref{fig:imodel}(d)]. Such a rationale also 
provides a natural framework for explaining the higher hole densities recently 
achieved under modified growth conditions.\cite{namkoong_APL2008,brochen_APL2013}

\begin{acknowledgements}
We acknowledge fruitful interactions with N. Grandjean. Financial support is 
acknowledged from the Swiss National Science Foundation (Grants Nos.\ 200020-152799).
We used computational resources of CSCS and CSEA-EPFL.
\end{acknowledgements}

\end{document}